\documentclass{article}
\usepackage{amssymb}

\usepackage{graphicx} 
\usepackage{amsmath}
\usepackage[utf8]{inputenc} 
\usepackage[top=1in, bottom=1in, left=0.8in, right=0.8in]{geometry}
\usepackage{amssymb}
\DeclareUnicodeCharacter{211D}{$\mathbb{R}$}
\usepackage[
	backend=biber,
	style=nature,
	sorting=nyt,
	maxnames=3,
	minnames=3,
	date=year,
	isbn=false,
	]{biblatex}
\usepackage{comment}
\usepackage{pdfpages}

\title{Classification of Solutions with Polynomial Energy Growth for 
the ${\mathrm SU}(n+1)$ Toda System on the Punctured Complex Plane}

\author{Genan Zhao}
\date{}

\begin{document}

\maketitle
\begin{abstract}
This paper investigates the classification of solutions satisfying the polynomial energy growth condition near both the origin and infinity to the ${\mathrm SU}(n+1)$ Toda system on the punctured complex plane $\mathbb{C}^*$. The ${\mathrm SU}(n+1)$ Toda system is a class of nonlinear elliptic partial differential equations of second order with significant implications in integrable systems, quantum field theory, and differential geometry.  Building on the work of A. Eremenko (J. Math. Phys. Anal. Geom., Volume 3 p.39-46), Jingyu Mu's  thesis, and others, we obtain the classification of such solutions by leveraging techniques from the Nevanlinna theory. In particular, we prove that the unitary curve corresponding to a solution with polynomial energy growth to the ${\mathrm SU}(n+1)$ Toda system on $\mathbb{C}^*$ gives a set of fundamental solutions to a linear homogeneous ODE of $(n+1)^{th}$ order, and each coefficient of the ODE can be written as a sum of a polynomial in $z$ and another one in $\frac{1}{z}$.

\end{abstract}

\newpage
\section*{Acknowledgement}
I would like to express my deepest gratitude to Professor Bin Xu at The University of Science and Technology of China for his pivotal role in introducing me to this intriguing topic, guiding me meticulously through the intricate process, and providing me with invaluable references that significantly enriched my research. My fascination with mathematical problems that have a robust physical underpinning has been a long-standing one, and this particular project allowed me to dive deeply into such complex issues. From the outset, my advisor Professor Xu showed immense support for my chosen research theme, recognizing its potential to contribute meaningfully to the field.

Throughout the course of this research, I have not only acquired a more profound comprehension of problems that merge the realms of mathematics and physics but also honed my skills in rigorous academic research. Under Professor Xu's expert guidance, I learned to navigate various academic challenges, from identifying the most relevant literature to applying complex theoretical concepts to solve practical problems. The skills I developed during this period include advanced analytical thinking, precise mathematical modeling, and the ability to synthesize information from disparate sources into a coherent and compelling argument.

Moreover, this journey has been about personal growth as much as it has been about academic development. I have emerged from this experience with a clearer vision of my academic and professional future, and a renewed commitment to contributing to the field of mathematical sciences. I am profoundly grateful to my advisor for his patience, wisdom, and encouragement throughout this challenging yet rewarding process.

In addition to my advisor's support, I must acknowledge the unwavering support of my family, particularly my parents. Their belief in my abilities and their constant encouragement have been the bedrock of my resilience and persistence. They have celebrated my successes and supported me through challenges, always encouraging me to pursue my passion for science and knowledge. Their emotional and moral support has been indispensable, and it has played a crucial role in my academic journey.

To sum up, the support and guidance I received from both my advisor and my family have been instrumental in the completion of this research. Their combined support not only fostered my academic skills but also reinforced my personal growth and professional aspirations. I am eternally grateful for their contributions to my journey.

\tableofcontents

\newpage

\section{Introduction}

The ${\mathrm SU}(n+1)$ Toda framework signifies a collection of non-linear elliptic differential systems of second order appearing in diverse math and physical scenarios, such as solvable systems, quantum field theory, and differential geometry. Studies into these frameworks, especially in bi-dimensional environments, have yielded profound understanding regarding the configuration and behavior of answers to intricate non-linear equations. Notably, the analysis of polynomial energy growth solutions for the ${\mathrm SU}(n+1)$ Toda scheme on the perforated complex field has attracted substantial interest because of its consequences in both theoretical and practical math domains.

The ${\mathrm SU}(n+1)$ Toda system can be formalized by the following collection of second order semi-linear elliptic partial differential equations:
\begin{equation}
(u_i)_{z\bar{z}} + \sum_{j=1}^n a_{ij} e^{u_j} = 0 \quad \text{in } \Omega \subseteq \mathbb{C}
\end{equation}
Here, \(i = 1, 2, \dots, n\), and \(\Omega\) is a domain within the complex plane $\mathbb{C}$; \((a_{ij})_{n \times n}\) denotes the Cartan matrix:

\begin{equation}
(a_{ij}) = 
\begin{pmatrix}
2 & -1 & 0 & \cdots & 0 \\
-1 & 2 & -1 & 0 & \cdots \\
0 & -1 & 2 & -1 & \cdots \\
\vdots & \vdots & \vdots & \ddots & \vdots \\
0 & \cdots & 0 & -1 & 2
\end{pmatrix}.
\end{equation}
This system is completely integrable as a set of partial differential equations in the sense that there exists a correspondence between its solutions and unitary curves on $\Omega$ (see subsection 2.2. for the details). Moreover, the system extends the well-known Liouville equation\cite{liouville1853equation} when \(n = 1\), reducing the equations to a simpler form:
\begin{equation}
-\Delta u_1=8 e^{u_1}
\end{equation}

Lately, considerable advancements have been achieved in elucidating the solutions of the ${\mathrm SU}(n+1)$ Toda framework. For instance, Jost and Wang\cite{jost2002classification} successfully delineated solutions with finite energy on \(\mathbb{C}\), while Eremenko \cite{eremenko2004toda} broadened these findings by employing the Nevanlinna theory\cite{nevanlinna1924klasse} to scrutinize solutions with polynomial energy growth at $\infty$ on $\mathbb{C}$.  Mu's thesis \cite{mu2024classification} offered a taxonomy of the Toda system's solutions with finite energy on \(\mathbb{C}^*\).

This study endeavors to augment prior findings by categorizing answers exhibiting polynomial energy growth for the $\operatorname{SU}(\mathrm{n}+1)$ Toda framework on the perforated complex field, utilizing Nevanlinna's theorem for meromorphic functions. Represented as $\mathbb{C}^*$, the perforated complex field introduces additional complexities due to the two singularities of $0$ and $\infty$. These two singularities affect the behavior of solutions and demand careful consideration for a comprehensive taxonomy.
This report explores the taxonomy of solutions for the ${\mathrm SU}(n+1)$ Toda framework established on the perforated complex field $\mathbb{C}^*$, centering on solutions wherein the energy of solutions displays polynomial growth  near both 0 and $\infty$. Specifically, our goal is to identify solutions $u=\left(u_1, \ldots, u_n\right)$ to the ${\mathrm SU}(n+1)$ system on $\mathbb{C}^*$ that satisfy the ensuing criterion:
\begin{equation}
\text{there exists}\  k > 0 \text{ such that } \frac{\sqrt{-1}}{2} \int_{\frac{1}{R}<|z|<R} e^{u_1} = O\left(R^k\right) \text{ as } R \to +\infty.
\end{equation}
Our key findings are encapsulated in the following two theorems:

\subsubsection*{Theorem 1}
{\textit
Let $\big(z^{b_0} \psi_0(z) \cdots z^{b_n} \psi_n(z)\big)$ represent a unitary curve associated corresponding to a solution $u=\left(u_1, \ldots, u_n\right)$ with polynomial energy growth to the ${\mathrm SU}(n+1)$ Toda system on $\mathbb{C}^*$, where $\psi_0, \ldots, \psi_n$ are holomorphic functions on $\mathrm{C}^*$. Then these $(n+1)$ holomorphic functions must have finite local growth order at both 0 and $\infty$.}

\subsubsection*{Theorem 2}
{\textit 
We use the notations in Theorem 1. 
$z^{b_0} \psi_0(z) \cdots z^{b_n} \psi_n(z)$ constitute a set of fundamental solutions set to the following homogeneous linear differential equation of order $(n+1)$ on $\mathbb{C}^*$:
\begin{equation}
y^{(n+1)} + \sum_{k=0}^{n-1} Z_{k+1}(z) y^{(k)} = 0,
\end{equation}
where the coefficients $Z_{k+1}(z)$ can be expressed as $P_k(z)+Q_k\left(\frac{1}{z}\right)$, with $P_k(\cdot)$ and $Q_k\left(\cdot\right)$ being polynomials.}

\noindent {\textbf Remark} We have identified that the coefficient functions have form $Z_{k+1}(\zeta)=P_k(\zeta)+Q_k\left(\frac{1}{\zeta}\right)$. However, whether these functions ensure that the monodromy matrix of some set of fundamental solutions to the preceding ODE lies within $\mathrm{SU}(\mathrm{n}+1)$ requires further investigation. We anticipate that additional conditions may need to be imposed, though a complete proof is yet to be developed.

The subsequent two sections of this manuscipt is structured as follows: Section 2  presents fundamental concepts regarding unitary curves and offer an in-depth discussion on how the Toda system relates to these curves; this section  also establishes the notation employed in Section 3. Section 3 is dedicated to the thorough proofs of Theorems 1 and 2.

\section{Preliminaries}
\subsection{${\mathrm SU}(n+1)$ Toda System}
\par We consider the following ${\mathrm SU}(n+1)$ Toda System in an arbitrary planar region $\Omega$
\begin{equation}
    (u_i)_{zz}+\sum_{j=1}^{n} a_{ij} e^{u_j}=0 \quad i=1,\ldots, n, 
\end{equation}
where
\begin{equation}
(a_{ij}) = 
\begin{pmatrix}
2 & -1 & 0 & \cdots & 0 \\
-1 & 2 & -1 & 0 & \cdots \\
0 & -1 & 2 & -1 & \cdots \\
\vdots & \vdots & \vdots & \ddots & \vdots \\
0 & \cdots & 0 & -1 & 2
\end{pmatrix}.
\end{equation}
The solution to the $\mathrm{SU}(\mathrm{n}+1)$ Toda system can be represented as $u=\left(u_1, \ldots, u_n\right)$, with the real-valued smooth functions $u_1, \ldots, u_n$ serving as the components.
The Toda system has its origins in problems from physics and is deeply intertwined with Lie algebra, complex analysis, harmonic mappings, and various other branches of mathematics. In the context of physics, it is often necessary that the Toda system satisfies specific energy restriction conditions. The polynomial energy growth condition can be summarized as:

\begin{equation}
\exists k > 0 \quad \text{s.t.} \quad \int_{\frac{1}{R} < |z| < R} e^{u_1} = O(R^{k}) \quad \text{as} \quad R \rightarrow +\infty
\end{equation}


\subsection{Correspondence between solutions and unitary curves}

In this subsection, we review the correspondence between solutions of the Toda system and unitary curves \cite[Section 2]{MR4741251}. We initiate by delineating a projective holomorphic curve and then illustrate how the solutions to the Toda system align with these curves.

A projective holomorphic curve $f: \Omega \to \mathbb{P}^n$ is described as a multi-valued holomorphic mapping that is non-degenerate, equipped with its monodromy representation $\mathrm{M}_f$: $\pi_1(\Omega, B) \to \operatorname{PSL}(n+1, \mathrm{C})$ functioning as a group homomorphism. This curve $f$ is designated {\textit unitary} if its monodromy belongs to $\operatorname{PSU}(n+1)$. Furthermore, a curve is considered totally unramified if each germ $f_z$ of $f$ is totally unramified at each location $z \in \Omega$, i.e. its Wronskian equals $1$ identically near $z$.

A totally unramified unitary curve $f:\Omega\to \mathbb{P}^n$ defines a solution to the ${\mathrm SU}(n+1)$ Toda system on $\Omega$ via the infinitesimal Pl\"ucker formula \cite[Lemma 2.2.]{MR4741251}.  We further explain how such a solution gives the corresponding totally unramified unitary curves, referencing Lemma 2.2 from Mu's thesis\cite{mu2024classification}.

Let $\left(u_1, u_2, \ldots, u_n\right)$ constitute a solution to the Toda system. Accordingly, there exists a function

$$
\Phi: \Omega \rightarrow {\mathrm SU}(n+1)
$$

such that the following relations hold:

\begin{align}
& \Phi^{-1} \cdot \Phi_z=\mathrm{U} \\
& \Phi^{-1} \cdot \Phi_{\bar{z}}=\mathrm{V}
\end{align}

This function $\Phi$ is known as the Toda mapping. If $\Omega$ is not simply connected, then $\Phi$ adopts a multi-valued nature.

Starting from the Toda mapping, we can construct a set of harmonic mappings:

\[
\left(f_0, f_1, \ldots, f_n\right) = \Phi \cdot \begin{pmatrix}
e^{w_0} & & & \\
& e^{w_1} & & \\
& & \ddots & \\
& & & e^{w_n}
\end{pmatrix}
\]

Since \( \Phi \) maps into \( {\mathrm SU}(n+1) \), it is evident that the vectors \( f_0, f_1, \ldots, f_n \) are mutually orthogonal in \( \mathrm{C}^{n+1} \), with:

\[
\left\|\hat{f}_i\right\| = e^{w_i}, \quad i=0, 1, \ldots, n.
\]

Using previous equations, we deduce that each \( \hat{f}_i \) satisfies the following relations:

\begin{align}
& \frac{\partial f_k}{\partial z} = f_{k+1} + \left(\log \left\|\hat{f}_k\right\|^2\right)_z \cdot f_k, \quad k=0, 1, \ldots, n-1, \\
& \frac{\partial \hat{f}_n}{\partial z} = \left(\log \left\|\hat{f}_n\right\|^2\right)_z \cdot \hat{f}_n, \\
& \frac{\partial f_k}{\partial z} = -\frac{\left\|\hat{f}_k\right\|^2}{\left\|\hat{f}_{k-1}\right\|^2} \hat{f}_{k-1}^{\perp}, \quad k=1, 2, \ldots, n, \\
& \frac{\partial f_0}{\partial z} = 0.
\end{align}

We therefore conclude that \( \hat{f}_0^T \) is a holomorphic curve. Moreover, for any \( k = 0, 1, \ldots, n \), we have :

$$
\Lambda_k\left(f_0^T\right)=f_0^T \wedge f_1^T \wedge \cdots \wedge f_k^T
$$

Therefore we get:

\begin{equation}
\left\|\Lambda_k\left(f_0^T\right)\right\|=\left\|f_0\right\|\left\|f_1\right\| \cdots\left\|f_k\right\|=e^{w_0+w_1+\cdots+w_k}
\end{equation}

where $k=0,1, \cdots, n$, in other words, we have:

\[
\|\hat{f_k}\| =
\begin{cases}
\frac{\|\Lambda_k(f_0^T)\|}{\|\Lambda_{k-1}(f_0^T)\|}, & k = 1, 2, \dots, n; \\
\|\Lambda_0(f_0^T)\|, & k = 0.
\end{cases} \quad \text{(2.22)}
\]

Specifically, we have:

\[
\Lambda_n(f_0^T) \equiv e_0 \wedge e_1 \wedge \dots \wedge e_n.
\]

Moreover, \(\hat{f_0^T}\) is also multivalued, like \(\Phi\), and because \(\Phi\) itself takes values in \({\mathrm SU}(n+1)\), the monodromy group of the multivalued curve \(\hat{f_0^T}\) is also a subgroup of \({\mathrm SU}(n+1)\). Therefore, \(\hat{f_0^T}\) is a unitary curve that satisfies the normalization conditions. Finally, we have:
\begin{equation}
\log \left\|\hat{f}_i\right\|= \begin{cases}\log \frac{\left\|\Lambda_1\left(f_0^T\right)\right\|^2}{\left\|\Lambda_0\left(f_T^T\right)\right\|^4}, & i=1 \\ \log \frac{\left\|\Lambda_i\left(f_0^T\right)\right\|^2\left\|\Lambda_{i-2}\left(f_0^T\right)\right\|^4}{\left\|\Lambda_{i-1}\left(f_0^T\right)\right\|^4}, & i=2,3, \cdots, n-1 \\ \log \frac{\| \Lambda_n\left(f_0^T \|^2\right.}{\left\|\Lambda_{n-1}\left(f_0^T\right)\right\|^4}, & i=n .\end{cases}
\end{equation}

This means that going from a solution $(u_1,..,u_n)$ of the Toda system, we can get a totally unramified unitary curve $\hat{f}_0^T$ on $\Omega$. And the solution of Toda system that corresponds to $\hat{f_0^T}$ is $(u_1,..,u_n)$. This establishes a clear correspondence between the Toda system's solution and unitary curves in the domain $D$.

\subsection{The Nevanlinna Theory}
The Nevanlinna theory stands as a fundamental aspect of complex analysis, offering significant insights into the behavior of meromorphic functions. This framework is essential for understanding how often and in what ways a meromorphic function assumes various values.

A meromorphic function $f(z)$ is depicted as a function that sustains analyticity throughout the entire complex field $\mathbb{C}$, excluding a limited count of isolated singular points. The expansion patterns of such functions can be articulated in relation to their order and the Nevanlinna characteristic function.

\textbf{Order of a Meromorphic Function:} The order $\rho(f)$ of a meromorphic function $f(z)$ is defined by:
\begin{equation}
\rho(f) = \limsup_{r \rightarrow \infty} \frac{\log T(r, f)}{\log r}
\end{equation}
where $T(r, f)$ denotes the Nevanlinna characteristic function, reflecting the rate at which $f(z)$ escalates as $|z|$ enlarges.

\textbf{Nevanlinna Characteristic Function:} This characteristic function, represented by \(T(r, f)\), acts as a gauge for the growth and value distribution of \(f(z)\) and is described by:

\begin{equation}
T(r, f) = m(r, f) + N(r, f)
\end{equation}

where \(m(r, f)\) is the Proximity Function, measuring the average closeness of \(f(z)\) to infinity within a circle of radius \(r\):
$$
m(r, f) = \frac{1}{2\pi} \int_0^{2\pi} \log^{+}\left|f\left(r e^{i\theta}\right)\right| d\theta
$$
Here, \(\log^{+} x = \max(\log x, 0)\).

\(N(r, f)\) is the Counting Function, tallying the number of poles of \(f(z)\) within the circle of radius \(r\):

$$
N(r, f) = \sum_{|a|<r} \log \frac{r}{|a|}
$$

where \(a\) represents the poles of \(f(z)\).

With the fundamentals of Nevanlinna theory in place, we can proceed to examine its core theorems, which offer essential insights into the behavior and distribution of values for meromorphic functions. These theorems are key instruments for investigating how these functions grow and how their values are distributed.

\textbf{First Main Theorem}: For a meromorphic function $f(z)$ and a complex constant $a$, the following holds:
\begin{equation}
T(r, f)=m\left(r, \frac{1}{f-a}\right)+N\left(r, \frac{1}{f-a}\right)+O(1)
\end{equation}
This theorem defines a connection between the growth rate of $f(z)$ and its value distribution characteristics.

\textbf{Second Main Theorem}: For distinct complex values \(a_1, a_2, \ldots, a_q\), the ensuing inequality holds true:
\begin{equation}
T(r, f) \geq \sum_{j=1}^{q} N\left(r, \frac{1}{f-a_j}\right)-(q-1)T(r, f)+S(r, f)
\end{equation}
where \(S(r, f)\) signifies a minor error term relative to \(T(r, f)\). This principle elucidates the value distribution of \(f(z)\) comprehensively.

Delving further into Nevanlinna theory's utilization within our investigation, we analyze outcomes from the paper \cite{KK2005} by Kondratyuk-Khrystyanyn. We focus on a meromorphic function $f$ charted on the annulus $A=\left\{z: \frac{1}{R_0}<|z|<R_0\right\}$, with $1<R_0 \leq+\infty$. The characteristic function $T_0(R, f)$ is established as:

$$
T_0(R, f)=\frac{1}{2\pi} \int_0^{2\pi} N_0\left(R, \frac{1}{f-e^{i\theta}}\right) d\theta
$$

for \(1<R<R_0\). Here, \(T_0(R, f)=m_0(R, f)-2 m(1, f)+N_0(R, f)\), where \(1<R<R_0\).

Define:

\begin{align}
y& N_1\left(R, \frac{1}{f-a}\right)=\int_{1 / R}^{1} \frac{n_1\left(t, \frac{1}{f-a}\right)}{t} dt \\
& N_2\left(R, \frac{1}{f-a}\right)=\int_{1}^{R} \frac{n_2\left(t, \frac{1}{f-a}\right)}{t} dt
\end{align}

where \(n_1\left(t, \frac{1}{f-a}\right)\) enumerates the poles of \(\frac{1}{f(z)-a}\) in \(\{z: t<|z| \leq 1\}\) and \(n_2\left(t, \frac{1}{f-a}\right)\) tallies the poles in \(\{z: 1<|z| \leq t\}\). Consequently:

$$
N_0\left(R, \frac{1}{f-a}\right)=N_1\left(R, \frac{1}{f-a}\right)+N_2\left(R, \frac{1}{f-a}\right)
$$

and:

\begin{equation}
N_0(R, f)=N_1(R, \infty)+N_2(R, \infty)
\end{equation}

This formulation broadens Nevanlinna theory's application by contemplating \(f\)'s behavior in annular zones.

\section{Proof and Property Analysis of The Toda System Solutions}

\subsection{Proof of Theorem 1}
Given that $z^{b_0} \psi_0(z) \cdots z^{b_n} \psi_n(z)$ is an entirely unramified and fulfills the normalized condition of a unitary curve associated with a solution $u=\left(u_1, \ldots, u_n\right)$ of the Toda system, where $\psi_0, \ldots, \psi_n$ are holomorphic functions on $\mathbb{C}^*$ and exhibit finite local growth order at both 0 and $\infty$.

We will first prove that the meromorphic functions $\frac{\psi_1}{\psi_0}\ldots\frac{\psi_n}{\psi_0}$: $\mathbb{C}^*\longrightarrow\mathbb{P}^1=\mathbb{C}\cup\{\infty\}$ have finite local growth order at both $0\textbf{ and }\infty$, and then that 
the holomorphic functions $\psi_0,...,\psi_n$: $\mathbb{C}^*\longrightarrow \mathbb{C}$ have the same property. 

\subsubsection*{Proposition 3.3}
{\textit For all $1\leq j \leq n$, $\frac{\psi_1}{\psi_0}\ldots\frac{\psi_n}{\psi_0}$ has finite local growth order at 0 and $\infty$.}

\noindent{\textbf Proof.} The proof is inspired by \cite[Chapter 3]{mu2024classification}. 
By using the infinitesimal Pl\"ucker formula, we could rewrite the polynomial energy growth condition for solutions in $\mathbb{C}$ by the following condition: there exists $k>0$ such that
\begin{equation}
\int_{\frac{1}{R}\leq|z|\leq R} f^*\omega_{\mathrm FS}
=O(R^k)
\end{equation}
as $R\to +\infty$, where $\omega_{\mathrm FS}$ means the Fubini-Study metric on $\mathbb{P}^n$.
Then we have:
\begin{equation}
    \int_{\frac{1}{R}\leq |z| \leq 1}f^*\omega_{\mathrm FS}\textit{, } \int_{1\leq |z| \leq R}f^*\omega_{\mathrm FS} = O(R^k)
\end{equation}

The order at \(\infty\) for a meromorphic function \( g: \mathbb{C} \rightarrow \mathbb{P}^1 \) is described by:
\begin{equation}
\limsup_{x \to \infty} \frac{\log(T(f, R))}{\log(R)}
\end{equation}
where \( T(f, R) \) denotes the Nevanlinna characteristic function.

Define $f(z)$ as $f(z)=(z^{b_0}\psi_0(z),..,z^{b_n}\psi_n(z)): \mathbb{C}^*\longrightarrow \mathbb{P}^n$. Let $B=(B_0,...B_n)\in \mathbb{C}^{n+1}$, $\|B\|=1$. 

Using these two functions, we now define a single-valued function $u_B$:
\begin{equation}
    u_B(\zeta)=\log{\frac{\|(z^{b_0}\psi_0,\ldots,z^{b_n}\psi_n)\|}{\|\sum_{j=0}^{n}B_j\psi_j\|}}
\end{equation}

Take annulus $\tilde{V}[R]=\{z|e^{-R}\leq|z|\leq e^R\}$, $\tilde{V}[R]\in \mathbb{C}^*$, and 
$$\tilde{V}[R]=V_1[R]\bigcup V_2[R]:=\{e^{-R}\leq |z|\leq 1\}\bigcup \{1\leq |z|\leq e^R\}.$$
On $\tilde{V}[R]$, because of $f(z)$ and $B$ being nondegenerate, the holomorphic function $\sum_{j=0}^{n}B_j\psi_j$ does not always equals to zero. Therefore apart from at most countable number of $R$, $u_B(\zeta)$ is defined. 

We can assume that $\sum_{j=0}^{n}B_j\psi_j$ never equals to zero on the boundary of $V_1[R]$. Taking
\[
d^c = \sqrt{-1}(\bar{\partial} - \partial),
\]
we clearly have \(d d^c = 2\sqrt{-1} \partial \bar{\partial}\). Notice,
\[
\frac{\partial^2}{\partial z \partial \bar{z}} \log |z - z_0| = \frac{\pi}{2} \delta(z - z_0),
\]
so apart from the zeros of  \(\sum_{j=0}^{n}B_j\psi_j\) , we have
\[
\frac{1}{2} d d^c u_B = \frac{1}{2} d d^c \log \|v\| = v^* \omega_{FS};
\]
Now, consider an \( R \) such that \(\tau = -e^R\) on which \(\sum_{j=0}^{n}B_j\psi_j \neq 0\). Let the zeros of \(\sum_{j=0}^{n}B_j\psi_j\) on \( V_1[R] \) be
\[
    z_1, z_2, \ldots, z_m
\]
Define
\[
U_j := \{z \mid |z - z_j| < \epsilon\}, \quad j = 1, 2, \ldots, m
\]
as small disks around the zeros, and we can write
\begin{equation}
\sum_{j=0}^{n}B_j\psi_j = (z - z_j)^{m_j} h_j(z)
\end{equation}
where \( h_j|_{U_j} \) is a non-zero holomorphic function. Finally, in coordinate chart 
$(\tau, \theta)=(\ln\,|z|, {\mathrm arg}\, z)$, we have
\begin{equation}
d^c = \frac{\partial}{\partial \tau} d\theta - \frac{\partial}{\partial \theta} d\tau.
\end{equation}
If on \( U_j \) we set
\[
z = z_j + \rho e^{\sqrt{-1} \varphi},
\]
then
\[
d^c = \rho \frac{\partial}{\partial \rho} d\varphi - \frac{1}{\rho} \frac{\partial}{\partial \varphi} d\rho.
\]
Combining the above, we can calculate:
\begin{align}
\frac{1}{\pi} \int_{V_1[R]} v^* \omega_{FS} &= \frac{1}{\pi} \lim_{\epsilon \to 0} \int_{V_1[R] - U_1 - U_2 - \cdots - U_m} v^* \omega\nonumber\\
&= \frac{1}{\pi} \lim_{\epsilon \to 0} \int_{V_1[R] - U_1 - U_2 - \cdots - U_m} \frac{1}{2} d d^c u_B \quad \nonumber\\
&= \frac{1}{2\pi} \lim_{\epsilon \to 0} \left( \int_{\tau = 0} d^c u_B - \int_{\tau = -R} d^c u_B - \sum_{i=1}^{m} \int_{\partial U_j} d^c u_B \right).
\end{align} 

In the above equations, we take the counterclockwise orientation as positive, and we can proceed further to obtain
\[
\int_{\tau = -R} d^c u_B = \int_{\tau = -R} \frac{\partial}{\partial \tau} u_B(\tau, \theta) d\theta 
= -\frac{d}{dR} \int_0^{2\pi} u_B(-R, \theta) d\theta;
\]
and also
\[
\lim_{\epsilon \to 0} \int_{\partial U_j} d^c u_B = \lim_{\rho \to 0} \int_0^{2\pi} \rho \frac{\partial}{\partial \rho} \left( \log \|(z^{b_0}\psi_0 \cdots z^{b_n}\psi_n)\| - \log \rho^{m_j} - \log h_j(z_j + \rho e^{\sqrt{-1}\varphi}) \right) d\varphi 
= -2\pi m_j.
\]
Also, define
\[
\frac{1}{2\pi} \int_{\tau = 0} d^c u_B = A_B,
\]
and 
\begin{center}
    $\tilde{n}_1(\rho, B)= \#\{z\in V_1[\rho]|\sum_{j=0}^n\, B_j\psi_j(z)=0\}$\\
     $\tilde{n}_2(\rho, B)= \#\{z\in V_2[\rho]|\sum_{j=0}^n\, B_j\psi_j(z)=0\}$
\end{center}

We will first look at $\tilde{n}_1$, and then substituting into equation \((22)\), we obtain
\begin{align}
\tilde{n}_1(\rho, B) + \frac{1}{2\pi} \frac{d}{dR} \int_0^{2\pi} u_B(-R, \theta) d\theta = \frac{1}{\pi} \int_{V_1[\rho]} v^* \omega - A_B, \quad
\label{3.14}
\end{align}
where $\tilde{n}_1(\rho, B)$ is in fact the number of zeros of the function \(\sum_{j=0}^{n}B_j\psi_j\) within the annulus \(V_1[\rho] := \{z \mid e^{-\rho} \leq \tau \leq 1\}\), which also represents the intersection number (counting multiplicities) of the holomorphic curve \(v_s(z) = [(\psi_0, \psi_1, \cdots, \psi_n)]\) with the hyperplane \(B^\perp \in Gr_n(n+1)\) perpendicular to $B$.

Note that equation \((23)\) is valid only when \(\sum_{j=0}^{n}B_j\psi_j|_{\tau = -R} \neq 0\). Now suppose there exist \(0 < 
\rho_1 < \rho_2\) such that \(u_B|_{-\rho_2 \leq \tau \leq -\rho_1}\) is well-defined, then integrating equation (\ref{3.14}) over \(\rho\), we have:
\begin{align}
\int_{\rho_1}^{\rho_2} \tilde{n}_1(\rho, B) d\rho + \frac{1}{2\pi} \int_{\rho_1}^{\rho_2} \frac{d}{d\rho} \left(\int_0^{2\pi} u_B(-\rho, \theta) d\theta\right) d\rho = \int_{\rho_1}^{\rho_2} \frac{1}{\pi} \left( \int_{V[e^\rho]} v^* \omega_{FS} - A_B \right) d\rho. \quad 
\label{3.15}
\end{align}

\noindent \textbf{Lemma 3.4.} (\cite{mu2024classification}) {\textit  The integral \(\int_{\rho_1}^{\rho_2} u_B(-\rho, \theta) d\rho\) is defined for any \( \rho_1 < \rho_2 \) and is continuous with respect to \(\rho\).}

Let \( V_1[\rho] \) contain a zero of $\sum_{j=0}^n B_j\psi_j$ \( z_0 = e^{-\rho_0 + \sqrt{-1}\theta_0} \), with multiplicity \( m_0 \), then following the previous notation, we can write
\[
\sum_{j=0}^{n}B_j\psi_j = (z - z_0)^{m_0} h_0(z),
\]
which leads to
\begin{center}
$u_B(-\rho, \theta) = \log \|(z^{b_0}\psi_0 \cdots z^{b_n}\psi_n)\| = \log \left| e^{-\rho+\sqrt{-1}\theta} - e^{-\rho_0+\sqrt{-1}\theta_0} \right| - m_0 \log \left| e^{-\rho_0+\sqrt{-1}\theta_0} \right| - \log h_0(-\rho, \theta).$
\end{center}
Therefore, if for every $\rho_0$, there exist $\int_{0}^{2\pi} \log \left| e^{-\rho+\sqrt{-1}\theta} - e^{-\rho_0+\sqrt{-1}\theta_0} \right| \, \text{d}\theta$ that is continuous at \(\rho = \rho_0\),  Lemma 3.4 is valid. We can summarize and prove by induction that
\[
I(\varrho) = \int_0^{2\pi} \log \left| \varrho e^{\sqrt{-1}\theta} - 1 \right| d\theta
\]
is continuous at \(\varrho = 1\). Here, \((\varrho, \theta)\) represents the traditional polar coordinates of \(z\). Notice,
\[
I(\varrho) = \int_0^{\pi} \log \left( 1 - 2\varrho \cos\theta + \varrho^2 \right) d\theta
\]
\[
= \begin{cases}
0, & \varrho \leq 1, \\
2\pi \log \varrho, & \varrho > 1.
\end{cases}
\]
Lemma 3.4 is therefore proven. 

The upper and lower bounds of the following integrals
$
\int_{\rho_1}^{\rho_2}\left(\tilde{n}_1, B\right) d \rho
$

\begin{equation}
\int_{\rho_1}^{\rho_2}\left(\frac{1}{\pi} \int_{V_1[\rho]} v^* \omega_{F S}-A_B\right) d \rho
\label{3.16}
\end{equation}
are continuous, so the interval \([\rho_1, \rho_2]\) can be expanded as long as \(u_B\) satisfies the condition \(u_B|_{-\rho_2 < \tau < -\rho_1}\). Then equation (\ref{3.15}) remains valid.

Now, fix any \(r > 0\). Suppose the function \(\sum_{j=0}^{n}B_j\psi_j\)  has zeros at \(\tau = r_1, r_2, \dots, r_k\) within the annulus \(V_1[r]\), where \(0 < r_1 < r_2 < \cdots < r_k \leq r\). Then we have:

\begin{align}
\tilde{n}_1(\rho, B) d\rho + \frac{1}{2\pi} \int_0^{2\pi} u_B(-\rho, \theta) d\theta \bigg|_{r_i}^{r_{i+1}} = \int_{r_i}^{r_{i+1}} \left( \frac{1}{\pi} \int_{V_1[\rho]} v^*\omega_{FS} - A_B \right) \rho, 
\label{3.17}
\end{align}

for \(i = 0, 1, \dots, k\). Here we set \(r_0 = 0\) and \(r_{k+1} = r\).

Adding the above \(k+1\) equations, we get:

\begin{align}
\int_0^r \tilde{n}_1(\rho, B) d\rho + \frac{1}{2\pi} \int_0^{2\pi} u_B(-\rho, \theta) d\theta \bigg|_0^r = \int_0^r \left( \frac{1}{\pi} \int_{V_1[\rho]} v^*\omega_{FS} - A_B \right) d\rho. 
\label{3.18}
\end{align}

However, the boundedness of our integrals implies that there is a uniform bound for \(\int_{V_1[\rho]} v^*\omega_{FS}\) over any \(\rho > 0\). Thus, we can set

\begin{align}
\frac{1}{\pi} \int_{V_1[\rho]} v^*\omega_{FS} - A_B \leq M_B. \quad
\label{3.19}
\end{align}

Furthermore, note that \(b_0, b_1, \dots, b_n \leq 0\). Thus, for any \(z \in D^*\), we have

\[
\|v\| = \left( \sum_{i=0}^n |z|^{2b_i} |\psi_i|^2 \right)^{\frac{1}{2}} \geq \left( \sum_{i=0}^n |\psi_i|^2 \right)^{\frac{1}{2}} = \|v_s\|,
\]

where \(\|v_s\|\) is defined similarly.
Furthermore, we have:

\[
\|\sum_{j=0}^{n}B_j\psi_j\| \leq \|(\psi_j)\| \|B\| = \|(\psi_j)\|.
\]

Thus, \(u_B\) and \(\int_0^{2\pi} u_B(-r, \theta) d\theta\) are always non-negative. Let

\[
\frac{1}{2\pi} \int_0^{2\pi} u_B(0, \theta) d\theta = C_B,
\]

Substituting equation (\ref{3.18}) and combining it with (\ref{3.19}), we get:
\begin{equation}
    \int_0^r \tilde{n}_1(\rho,B)d\rho=O(e^{kr})\text{ as } r\longrightarrow \infty
\end{equation}
Similarily, from $\tilde{n}_2(\rho, B)= \{z\in V_2[\rho]|\sum_{j=0}^n\, B_j\psi_j(z)=0\}$, we get:
\begin{equation}
    \int_0^r \tilde{n}_2(\rho,B)d\rho=O(e^{kr})\text{ as }r\longrightarrow \infty
\end{equation}

Therefore we have
\begin{equation}
    \int_0^r [\tilde{n}_1 (\rho,B)+\tilde{n}_2(\rho,B)]d\rho =O(e^{kr})
\end{equation}
Let
\begin{equation}
      n_1(R, B) = \#\{\frac{1}{R}\leq |z|\leq 1|\sum_{j=0}^n B_j\psi_j=0\}
\end{equation}
\begin{equation}    
    n_2(R, B) = \#\{1 \leq |z|\leq R|\sum_{j=0}^n B_j\psi_j=0\}
\end{equation}
Through change of variables, we obtain:
\begin{equation}
    O(e^{kr})=\int_0^r \tilde{n}_1 (\rho, B)d\rho=\int_0^{e^r}\frac{n_1(\rho,B)}{\rho}d\rho=N_1(e^r,B)
\end{equation}
Similarily, we have:
\begin{equation}
    O(e^{kr})=\int_0^{e^r}\frac{n_2(\rho,B)}{\rho}d\rho=N_2(e^r,B)
\end{equation}
Hence
\begin{center}
    $N_0(R,B)=N_1(R,B)+N_2(R,B)=O(R^k)\textit{ as }R\longrightarrow \infty$
\end{center}
Using the notations in \cite{KK2005}, we have:
\begin{equation}
    N_0\left(R,\frac{1}{\sum_{j=0}^n B_j\psi_j-0}\right)=O(R^k)
\end{equation}
Taking $B=(-\lambda, 0 \ldots 0,1,0 \ldots 0) / \sqrt{1+|\lambda|^2}$ where $\lambda \in \mathbb{C}$. We have
We have
\begin{equation}
    N_0\left(R,\frac{1}{\frac{\psi_j}{\psi_0}-\lambda}\right)=O(R^k)
\end{equation}
By Lemma 1 in \cite{KK2005}, for all $R>1$ we have
\begin{equation}
    T_0\left(R,\frac{\psi_j}{\psi_0}\right)=\frac{1}{2\pi}\int_0^{2\pi} N_0\left(R,\frac{1}{\frac{\psi_j}{\psi_0}-e^{i\theta}}\right){\mathrm d}\theta
\end{equation}
Hence $T_0(R,\frac{\psi_j}{\psi_0})=O(R^k)$, and $\frac{\psi_j}{\psi_0}$ has finite local growth order at both 0 and $\infty$. Q.E.D.\\

Theorem 1 follows from the following

\subsubsection*{Proposition 3.5} 
{\textit Let $\frac{\psi_j}{\psi_0}$ for $1 \leq j \leq n$ possess finite order at both 0 and $\infty$, and let $f(z)=\left(z^{b_0} \psi_0(z), \ldots, z^{b_n} \psi_n(z)\right)$ have Wronskian equal to $1$ identically on $\mathbb{C}^*$. Then, the functions $\psi_0, \ldots, \psi_n$ have finite local growth order at both 0 and $\infty$.}

\noindent {\textbf Proof.} 
Define a lifting of $f(z)$ as $\tilde{f}(z)=(1, \frac{f_1}{f_0},...\frac{f_n}{f_0})=(1,z^{b_1-b_0}\frac{\psi_1}{\psi_0},...)$
We have by \cite[(2.33)]{mu2024classification} that 
\begin{equation*}
\begin{gathered}
\Lambda_n(f)=e_0 \wedge e_1 \wedge \cdots \wedge e_{n+1} \\
=\mathrm{f}_0^{n+1} \Lambda_n\left(f_0^{n+1} z^{\sum\left(b_j-b_0\right)-\frac{n(n+1)}{2}} G_n\left(b_1-b_0, \ldots, b_n-b_0, \frac{\psi_1}{\psi_0} \cdots \frac{\psi_n}{\psi_0}\right)\right),
\end{gathered}
\end{equation*}
where $G_n=G_n\left(b_1-b_0, \ldots, b_n-b_0, \frac{\psi_1}{\psi_0}, \ldots, \frac{\psi_n}{\psi_0}\right)$ is a polynomial about the derivatives
of $\frac{\psi_1}{\psi_0}, \ldots, \frac{\psi_n}{\psi_0}$ up to order $n$.
Therefore $f_0=z^{b_0}\psi_0(z)=z^{\sum(b_i-b_0)-\frac{n(n+1)}{2}}G_n(b_1-b_0,\ldots,b_n-b_0,\frac{\psi_1}{\psi_0}\ldots\frac{\psi_n}{\psi_0})$

On the other hand, in subsection 2.2.4 of Steinmetz \cite{MR3676902}, it is proposed that the derivative $f^{(p)}$ possesses an associated pole counting function $N\left(r, f^{(p)}\right)=N(r, f)+p \bar{N}(r, f)$. By representing $\frac{f^{(p)}}{f}$ as a product of successive derivatives,

$$
\frac{f^{(p)}}{f}=\frac{f^{(p)}}{f^{(p-1)}} \cdot \frac{f^{(p-1)}}{f^{(p-2)}} \cdots \frac{f^{\prime}}{f}
$$

and acknowledging that

$$
f^{(p)}=\frac{f^{(p)}}{f} \cdot f
$$

it can be established that

$$
T\left(r, f^{(p)}\right) \leq T(r, f)+p \bar{N}(r, f)+S(r, f) \leq(p+1) T(r, f)+S(r, f)
$$
which delineates the boundaries on the growth of the function and its derivatives \cite{MR3676902}. Consequently, the elevated function $\tilde{f}(z)$, constructed from these functions, exhibits finite order.

Summing up the preceding two paragraph, we find that the function $G_n: \mathbb{C}^* \rightarrow \mathbb{P}^1$ has finite local growth order at both 0 and $\infty$, and the functions $\psi_0, \ldots, \psi_n$ display the same property. Q.E.D.

\subsection{Proof of Theorem 2}
We at first show the following statement: 
{\textit Let u=($u_1,...u_n$) be a solution with polynomial energy growth at both $0$ and $\infty$ to the ${\mathrm SU}(n+1)$ Toda system on $\mathbb{C}^*$. Then there exists a totally unramified unitary curve 
\begin{equation}
f=\left(f_0^{(z)}, \ldots, f_n^{(z)}\right)=\left(z^{b_0} \psi_0(z), \ldots,  z^{b_n} \psi_n(z)\right)
\end{equation}
corresponding to $u$, 
where $b_0...b_n\in \mathbb{R}$, $\psi_0...\psi_n$ are holomorphic functions in $\Omega$, such that the monodromy representation $\pi_1(\mathbb{C}^*, 1)\to {\mathrm SU}(n+1)$ of $f$ 
is diagonal.
Furthermore, the components $f_0, f_1, \ldots, f_n$ constitute a set of fundamental solutions to the following homogeneous linear differential equations  of order $(n+1)$ :
\begin{center}
    $y^{(n+1)}+\sum_{k=0}^{n-1}Z_{k+1}y^{(k)}=0$
\end{center}
where the coefficient functions $Z_k$ are holomorphic in $\mathbb{C}^*$.}

\noindent{\textbf Proof.}
Since $\pi_1(\mathbb{C}^*,1) \cong \mathbb{Z} =<[\gamma]> $ where $\gamma$ is a positive loop around the puncture near the origin. We could choose 
a totally unramified unitary curve $f=\mathbb{C}^* \longrightarrow \mathbb{P}^n$ corresponding to $u$ such that its monodromy has form $[\gamma]\mapsto {\mathrm diag}\left(e^{2\pi \sqrt{-1} b_0},\ldots, e^{2\pi \sqrt{-1} b_n}\right)$ with 
\begin{equation}
\label{mono}  
\prod_{i=0}^n\, e^{2\pi \sqrt{-1} b_i}=1
\end{equation}
and $b_i\in\mathbb{R}$. 
In particular, $f_j:\mathbb{C}^*\longrightarrow\mathbb{C}$ has monodromy $[\gamma]\mapsto e^{2\pi \sqrt{-1} b_j}$, where $j=0,1..,n$. 

Since we define $f$ as a totally unramified  curve, we can choose a homogeneous representation such that the wronskian of $f$ as:
\begin{equation}
\begin{vmatrix}

f_0 & f_1 & \cdots & f_n \\
f_0^{(1)} & f_1^{(1)} & \cdots & f_n{(1)} \\
\vdots & \vdots & \ddots & \vdots \\
f_0^{(n)} & f_1^{(n)} & \cdots & f_n^{(n)} \\
\end{vmatrix} \equiv e_0\wedge...\wedge e_n =1
\end{equation}

Since the Wronskian here does not vanish identically, it nowhere vanishes. Hence $(f_0,...f_n)$ is linearly independent. 
The ODE of (n+1)th order has the form of:
\begin{equation}
\begin{vmatrix}
y^{(n+1)} & y^{(n)} & \cdots & y \\
f_0^{(n+1)} & f_0^{(n)} & \cdots & f_0 \\
f_1^{(n+1)} & f_1^{(n)} & \cdots & f_1 \\
\vdots & \vdots & \ddots & \vdots \\
f_n^{(n+1)} & f_n^{(n)} & \cdots & f_n \\
\end{vmatrix} = 0
\end{equation}
By using (\ref{mono}), We have that the coefficient of $y^{(n+1)}$ equals 1 from the totally unramified property of
$f$. Moreover, the coefficient of $y^{(k)}$ for each $0\leq k\leq n$ is a single-valued holomorphic function on $\mathbb{C}$ by
$(ref{mono})$. At last,  utilizing the argument in \cite[Proposition 3.10]{MR1700776}, we could see the coefficient of the derivative of the second to highest order vanishes identically. Q.E.D. \\

We complete the proof of Theorem 2 by showing that
{\textit each $Z_{k+1}(z)$ has poles at both $0$ and $\infty$.}

Actually, by Proposition 3.5,   $\psi_0...\psi_n$ are holomorphic functions on $\mathbb{C}^*$ with finite local growth order at both $0$ and $\infty$.  Using Theorem 2 in B\"ohmer \cite{MR0296381}, we find that each holomorphic function $Z_{k+1}$ on $\mathbb{C}^*$ has poles at both $0$ and $\infty$. Hence, there exist two polynomials $P_k$ and $Q_k$ such that
\begin{equation}
  Z_{k+1}(z)=P_k(z)+Q_k(1/z),\quad z\in \mathbb{P}^1=\mathbb{C}\cup \{\infty\}.
\end{equation}

\nocite{*}

\printbibliography

@book{teschl2013,
  author    = {Gerald Teschl},
  title     = {Ordinary Differential Equations and Dynamical Systems},
  year      = {2013},
  publisher = {American Mathematical Society},
  address   = {Providence, RI},
  volume    = {140},
  series    = {Graduate Studies in Mathematics}
}

@article{KK2005,
  author = {A. Kondratyuk and A. Khrystyanyn},
  title = {An extension of Nevanlinna value distribution theory for meromorphic functions on annuli},
  journal = {Mathematical Studies},
  year = {2005}
}

@article{jost2002classification,
  title={Classification of solutions of a Toda system in ℝ 2},
  author={Jost, J{\"u}rgen and Wang, Guofang},
  journal={International Mathematics Research Notices},
  volume={2002},
  number={6},
  pages={277--290},
  year={2002},
  publisher={OUP}
}

@article {eremenko2004toda,
    AUTHOR = {Eremenko, A.},
     TITLE = {A {T}oda lattice in dimension 2 and {N}evanlinna theory},
   JOURNAL = {J. Math. Phys. Anal. Geom.},
  FJOURNAL = {Journal of Mathematical Physics, Analysis, Geometry},
    VOLUME = {3},
      YEAR = {2007},
    NUMBER = {1},
     PAGES = {39--46, 129},
      ISSN = {1812-9471,1817-5805},
   MRCLASS = {37K60 (30D35 37J35 37K20)},
  MRNUMBER = {2304085},
MRREVIEWER = {Iskander\ A.\ Taimanov},
}

@article{calabi1953metric,
  title={Metric Riemann surfaces},
  author={Calabi, E},
  journal={Ann. of Math. Studies},
  volume={30},
  pages={77--85},
  year={1953}
}

@article{chen1991classification,
  title={Classification of solutions of some nonlinear elliptic equations},
  author={Chen, Wenxiong and Li, Congming},
  year={1991}
}

@article{liouville1853equation,
title={Sur l’{\'e}quation aux diff{\'e}rences partielles $\frac{d^2\log\lambda}{du dv}\pm\frac{\lambda}{2a^2}=0.$},
author={Liouville, Joseph},
  journal={Journal de math{\'e}matiques pures et appliqu{\'e}es},
  volume={18},
  pages={71--72},
  year={1853}
}

@article{nevanlinna1924klasse,
  title={{\"u}ber eine Klasse meromorpher Funktionen},
  author={Nevanlinna, Rolf},
  journal={Mathematische Annalen},
  volume={92},
  number={3},
  pages={145--154},
  year={1924},
  publisher={Springer}
}

@book {MR3676902,
    AUTHOR = {Steinmetz, Norbert},
     TITLE = {Nevanlinna theory, normal families, and algebraic differential
              equations},
    SERIES = {Universitext},
 PUBLISHER = {Springer, Cham},
      YEAR = {2017},
     PAGES = {xviii+235},
      ISBN = {978-3-319-59799-7, 978-3-319-59800-0},
   MRCLASS = {30-01 (30D35 30D45 34A09 34M05 34M10)},
  MRNUMBER = {3676902},
MRREVIEWER = {Ilpo\ Laine},
       DOI = {10.1007/978-3-319-59800-0},
       URL = {https://doi.org/10.1007/978-3-319-59800-0},
}

@article {MR4741251,
    AUTHOR = {Mu, Jingyu and Shi, Yiqian and Sun, Tianyang and Xu, Bin},
     TITLE = {Classifying solutions of {${\rm SU}(n+1)$} {T}oda system
              around a singular source},
   JOURNAL = {Proc. Amer. Math. Soc.},
  FJOURNAL = {Proceedings of the American Mathematical Society},
    VOLUME = {152},
      YEAR = {2024},
    NUMBER = {6},
     PAGES = {2585--2600},
      ISSN = {0002-9939,1088-6826},
   MRCLASS = {37K10 (35J47)},
  MRNUMBER = {4741251},
       DOI = {10.1090/proc/16785},
       URL = {https://doi.org/10.1090/proc/16785},
}

@article {MR0296381,
    AUTHOR = {B\"ohmer, Klaus},
     TITLE = {\"Uber lineare {D}ifferentialgleichungen mit {L}\"osungen von
              endlicher {W}achstumsordnung},
   JOURNAL = {Arch. Math. (Basel)},
  FJOURNAL = {Archiv der Mathematik},
    VOLUME = {22},
      YEAR = {1971},
     PAGES = {394--400},
      ISSN = {0003-889X,1420-8938},
   MRCLASS = {34A20},
  MRNUMBER = {296381},
MRREVIEWER = {J.\ Morav\v c\'ik},
       DOI = {10.1007/BF01222594},
       URL = {https://doi.org/10.1007/BF01222594},
}

@book {MR1700776,
    AUTHOR = {Kohno, Mitsuhiko},
     TITLE = {Global analysis in linear differential equations},
    SERIES = {Mathematics and its Applications},
    VOLUME = {471},
 PUBLISHER = {Kluwer Academic Publishers, Dordrecht},
      YEAR = {1999},
     PAGES = {xvi+526},
      ISBN = {0-7923-5605-5},
   MRCLASS = {34M40 (33E30 34M35)},
  MRNUMBER = {1700776},
MRREVIEWER = {Donald\ A.\ Lutz},
       DOI = {10.1007/978-94-011-4605-0},
       URL = {https://doi.org/10.1007/978-94-011-4605-0},
}

@article {MR2150985,
    AUTHOR = {Khrystiyanyn, A. Ya. and Kondratyuk, A. A.},
     TITLE = {On the {N}evanlinna theory for meromorphic functions on
              annuli. {I}},
   JOURNAL = {Mat. Stud.},
  FJOURNAL = {Matematichn\=\i\ Stud\=\i\"i. Prats\=\i\ L\cprime v\=\i
              vs\cprime kogo Matematichnogo Tovaristva},
    VOLUME = {23},
      YEAR = {2005},
    NUMBER = {1},
     PAGES = {19--30},
      ISSN = {1027-4634,2411-0620},
   MRCLASS = {30D35},
  MRNUMBER = {2150985},
}

@phdthesis{mu2024classification, 
title = {The classification and construction of solutions with finite energy to {SU}(n+1) {T}oda system on Riemann surfaces}, author = {Jingyu Mu}, year = {2024}, school = {University of Science and Technology of China}, type = {PhD Dissertation}, address = {Hefei, China}, month = {May}, note = {Speciality: Pure Mathematics} }
\end{document}